\documentclass[10pt,final]{iopart}
\pdfoutput=1

\usepackage[T1]{fontenc}
\usepackage[english,UKenglish]{babel}
\usepackage[utf8]{inputenc}
\usepackage{latexsym}
\usepackage{graphicx}
\usepackage{color}   
\usepackage{amsfonts}
\usepackage{amssymb} 
\usepackage{amsthm}
\usepackage{bbm}     
\usepackage{ae}		
\usepackage{type1cm}	
\usepackage{mathrsfs}	
\usepackage{enumitem}	
\usepackage[sort&compress,numbers]{natbib}
\usepackage{color,soul}
\usepackage[small,bf]{caption}		

\usepackage[bookmarks,	
            bookmarksopen=false,
            bookmarksnumbered=true,
            pdfstartview=Fit, 	
	    linkbordercolor={0 0 1},		
            citebordercolor={0 0.80 0.2},	
	    urlbordercolor={0.55 0.11 0.38},	
	    colorlinks,			
            linkcolor=blue,	
	    citecolor=blue,	
            urlcolor=blue,	
	    filecolor=blue 		
	    pdftitle={Ordered Measurements of Permutationally-Symmetric Qubit Strings},
            pdfauthor={Alexander Hentschel},
	    plainpages=false,
	    pdfpagelabels]{hyperref}

\newcommand{\hs}[1]      {\hspace{#1}}
\newcommand{\vs}[1]      {\vspace{#1}}

\newcommand{\tn}[1]      {{\rm #1}}

\newcommand{\braket}[2]  {\langle #1 | #2 \rangle}
\newcommand{\bra}[1]     {\langle #1 |}
\newcommand{\ket}[1]     {| #1 \rangle}

\newcommand{\Pop}        {\mathcal{P}}

\newcommand{\PiStateFull}[2]    {\ket{#1}_{[#2]}^{}}

\newcommand{\Span}              {\mathop{\mathrm{span}}}

\newcommand{\eqref}[1]{(\ref{#1})}
\newcommand{\binom}[2]{\left({{#1}\atop{#2}}\right)}

\DeclareMathAlphabet{\mathitbf}{OML}{cmm}{b}{it}

\newtheoremstyle{definition}{\topsep}{\topsep}%
     {\itshape}
     {}
     {\bfseries}
     {}
     {\newline}
     {\thmname{#1}\thmnumber{ #2}:\thmnote{ \textnormal{#3}}}

\newcounter{PropositionCounter}

\theoremstyle{definition}
\newtheorem{definition}{Definition}
\newtheorem{remark}{Remark}
\newtheorem{lemma}[PropositionCounter]{Lemma}
\newtheorem{corollary}[PropositionCounter]{Corollary}
\newtheorem{theorem}[PropositionCounter]{Theorem}

\definecolor{darkred}{rgb}{0.80,0.0,0.0}
\definecolor{darkblue}{rgb}{0.0,0.0,0.55}
\definecolor{darkgreen}{rgb}{0.0,0.4,0.0}
\definecolor{maroon}{rgb}{0.40,0.08,0.30}

  \def\newblock{\hskip .11em plus .33em minus .07em}

\begin{document}

\addtolength{\topmargin}{-0.5cm}

\begin{center}
	 {\bf \large Ordered Measurements of Permutationally-Symmetric Qubit Strings} 
\end{center}

\begin{center}
  Alexander Hentschel and Barry C. Sanders
\end{center}

\begin{center}
  {\it Institute for Quantum Information Science, University of Calgary, Calgary, Alberta, Canada T2N 1N4}\\
\end{center}
\vspace{0.15in}	

\begin{center}
 \begin{minipage}{0.8\textwidth}
   \textbf{Abstract:}
    We show that any sequence of measurements on a permutationally-symmetric (pure or mixed) multi-qubit string leaves the unmeasured qubit substring also permutationally-symmetric. In addition, we show that the measurement probabilities for an arbitrary sequence of single-qubit measurements are independent of how many unmeasured qubits have been lost prior to the measurement. Our results are valuable for quantum information processing of indistinguishable particles by post-selection, e.g.\ in cases where the results of an experiment are discarded conditioned upon the occurrence of a given event such as particle loss. Furthermore, our results are important for the design of adaptive-measurement strategies, e.g.\ a series of measurements where for each measurement instance, the measurement basis is chosen depending on prior measurement results. 
 \end{minipage}
\end{center}
\vspace{0.0in} 	



\section{Introduction}

\noindent%
An $n$-partite quantum state is permutationally-symmetric (also referred to as `symmetric states' \cite{Browne&Plenio:PhysRevA.66.2002,Stockton&al:PhysRevA.67.2003,Bartlett&Spekkens:PRL2003}) 
if it is invariant under arbitrary permutations of the $n$ subsystems. Such `permutationally-symmetric states' are optimal for quantum phase metrology \cite{Quantum_phase_estimation_with_lossy_interferometers:PRA:2009}; specifically, highly-entangled permutationally-symmetric states are a key resource for adaptive phase measurement \cite{PhysRevLett.85.5098,PhysRevA.63.053804,Berry-Wiseman:PhysRevA:2009,QLearning:hentschel:PRL:2010,QLearning:hentschel:ITNG:2010}, as they enable precision measurements close to the fundamental Heisenberg limit in the ideal pure-state and unitary-evolution setting \cite{Caves-Carlton-1981-PhysRevD,1986:Yurke-Samuel-Klauder}. 
Adaptive measurement schemes use successive measurements where the type of measurement, e.g.\ the measurement basis, is chosen based on prior measurement results. For devising practically applicable adaptive-measurement schemes, we require a theory that is sufficiently general and includes experimentally-relevant cases such as loss. For example, if an $n$-partite quantum system is used for an adaptive measurement, but only $m$, with $m<n$, qubits arrive at the detector, the missing $n-m$ qubits are said to be lost. 
An adaptive-measurement scheme is said to be `loss-independent' if the measurement probabilities only depend on how many qubits are measured but not on how many and which of the unmeasured qubits have been lost. Even though the loss of qubits typically occurs randomly, we can study the case where the loss is fixed by discarding selected experimental outcomes; this is post-selection.
Here we study the robustness of permutationally-symmetric states in the presence of loss and show that permutationally-symmetric states are especially useful for adaptive quantum measurement strategies.

Although errors and losses do not typically preserve permutation symmetry, we show that any sequence of operations (such as measurements, errors or loss) that do not necessarily preserve permutation symmetry nonetheless leave the remaining untouched qubit substring permutationally-symmetric. 
Our focus is on adaptive-measurement schemes with permutationally-symmetric states as inputs. For example, consider an optical adaptive phase-estimation scheme, where individual photons of an entangled, permutationally-symmetric state are fed sequentially into an interferometer \cite{QLearning:hentschel:PRL:2010,PhysRevLett.85.5098}. 
After each output photon is measured, the measurement for the next input photon is adaptively changed based on the measurement history.
As each input photon of the input state is processed at a different time, most operations, such as measurement and quantum errors, are not permutationally-symmetric. 
Specifically we show that qubit loss does not harm adaptive-measurement schemes that use permutationally-symmetric states as input, in the sense that all measurement probabilities are loss-independent, despite the fact that losses raise the cost in terms of number of required photons and integration time for the measurement procedure.

Our paper is structured as follows. 
In section two, we present the mathematical formalism for permutationally-symmetric states from a quantum-information perspective.  
The third section deals with the effects of measurements and loss on permutationally-symmetric states. As our focus is on adaptive measurements, we specifically analyse the effects of sequential single-qubit measurements and loss on permutationally-symmetric states. In section four, we present implications for adaptive measurement schemes that utilise pure permutationally-symmetric input states and projective-valued measures to estimate parameters of a unitary single-qubit quantum channel. In the fifth section, we draw conclusions and summarise our work.


\section{Permutationally-symmetric qubit strings}

\noindent%
We begin with introducing some notation and basic concepts.
The Hilbert space for a single qubit is 
\begin{eqnarray} \label{eq:Definition_Qubit_Space}
	\mathscr{H}_2 = \Span \{ \ket{0},\ket{1} \}\,,
\end{eqnarray}
and a pure $n$-qubit state $\ket{\Psi_{n}}$ is 
\begin{eqnarray}
 \ket{\Psi_{n}} \in \mathscr{H}_2^{\otimes n} = \Span \big\{ \ket{\beta_n \beta_{n-1} \cdots \beta_1}\, : ~ \beta_n,\ldots, \beta_1 \in \{0,1\} \big\} 
\end{eqnarray}
for $\beta_i$ the binary value of the $i^\tn{\,th}$ qubit. 
As we restrict our attention to two-level quantum systems, the relevant Hilbert spaces are $\mathscr{H}_2 = \mathbbm{C}^2$ and tensor products spaces of $\mathbbm{C}^2$.
A mixed state of $n$ qubits is a unit-trace operator 
\begin{eqnarray}\label{eq:Definition_B(H_2^n)}
 \rho \in \mathcal{B}(\mathscr{H}_2^{\otimes n}) =  \Span\big\{ \ket{\mathitbf{a}}\bra{\mathitbf{b}}\, : \mathitbf{a},\mathitbf{b} \in \{0,1\}^n\big\},
\end{eqnarray}
where $\mathcal{B}(\mathscr{H})$ denotes the space of linear operators on the Hilbert space $\mathscr{H}$.

Let $S_{n}$ be the symmetric group, which is the group of all permutations of $n$ objects. In a qubit string of length $n$, each individual qubit has a unique position $i\in \{1,2,\ldots,n\}$. Here a permutation $\pi\in S_n$ is a bijective map 
\begin{eqnarray}
   \pi: \{1,2,\ldots,n\} \rightarrow \{1,2,\ldots,n\}\, ,
\end{eqnarray}
which reshuffles the positions of the qubits. In Definition \ref{def:permutation_Operator} below, we establish a unitary representation for the permutations.
\begin{definition}[Qubit-permutation operator]\label{def:permutation_Operator}
   For $\pi\in S_n$, the unitary representation of the qubit-permutation operator is 
   \begin{eqnarray}\fl
 	\quad 
		\mathcal{P}(\pi) : \mathscr{H}_2^{\otimes n} \rightarrow \mathscr{H}_2^{\otimes n}: 
		\ket{\beta_n \beta_{n-1} \cdots \beta_1} \mapsto 
				\mathcal{P}(\pi)\ket{\beta_n  \cdots \beta_1} = \ket{\beta_{\pi^{-1}(n)} \beta_{\pi^{-1}(n-1)} \cdots \beta_{\pi^{-1}(1)}}.
   \end{eqnarray}
\end{definition}

\begin{remark}\label{eq:decomposition_of_Permutation_operator}
   $\Pop$ is a unitary representation of $\mathcal{S}_n$. Therefore, $\Pop$ has the properties $\Pop(\pi)^\dagger = \Pop(\pi)^{-1} = \Pop(\pi^{-1})$ and 
   $\mathcal{P}(\pi_1 \circ \pi_2 ) = \mathcal{P}(\pi_1) \mathcal{P}(\pi_2 )$.
\end{remark}

\noindent%
The space $\mathscr{H}_2^{\otimes n}$ is composed of $n$ single-qubit Hilbert spaces, by default labelled $\mathscr{H}^{(n)}_2, \ldots, \mathscr{H}^{(2)}_2$ and $\mathscr{H}^{(1)}_2$. Most of the time, it is sufficient to use the compressed notation $\mathscr{H}_2^{\otimes n}$ since the actual labels are irrelevant.
However, if we, for instance, discard qubits in the string $\ket{010} \in \mathscr{H}_2^{\otimes 3}$, one has to explicitly label the Hilbert spaces of the remaining qubits in order to distinguish the outcomes for losing different qubits. Hence, if necessary, we write $\ket{010} \in \mathscr{H}^{(3)}_2 \otimes  \mathscr{H}^{(2)}_2 \otimes \mathscr{H}^{(1)}_2$ or simply $\ket{010} \in \bigotimes_{i\in \{3,2,1\}} \mathscr{H}^{(i)}_2$. 

The qubit-permutation operator $\mathcal{P}(\pi)$, with $\pi \in S_n$, acts on the $n$-qubit Hilbert space $\mathscr{H}_2^{\otimes n}$. However, $\mathscr{H}_2^{\otimes n}$ could be a subspace of a larger Hilbert space  $\mathscr{H}_2^{\otimes m}$ with $m>n$. Consider for instance a permutation $\pi \in S_3$, which reshuffles the position of 3 qubits. Given the state 
$\ket{\psi} = \ket{01001} \in \mathscr{H}_2^{\otimes 5}$, the permutation could be applied to any three of the five qubits.
To avoid ambiguity, we indicate, if necessary, the positions of the qubits, $\pi$ is acting on by a subscript on permutation operator, i.e.\ 
$\mathcal{P}_{_{\hs{-2pt}\mathscr{S}}}(\pi) : \bigotimes_{i\in\mathscr{S}} \mathscr{H}^{(i)}_2 \rightarrow \bigotimes_{i\in\mathscr{S}} \mathscr{H}^{(i)}_2$.
For convenience of notation, we extend the mapping $\mathcal{P}_{_{\hs{-2pt}\mathscr{S}}}(\pi)$ to any Hilbert space $\mathscr{H} = \bigotimes_{i\in \mathscr{S}'} \mathscr{H}^{(i)}_2$ with $\mathscr{S} \subseteq \mathscr{S}'$, via 
\begin{eqnarray}\fl
  \mathcal{P}_{_{\hs{-2pt}\mathscr{S}}}(\pi) \equiv \mathcal{P}_{_{\hs{-2pt}\mathscr{S}}}(\pi) \otimes \mathbbm{1} 
  ~\in          \mathcal{B}\Big(\bigotimes_{i\in \mathscr{S}} \mathscr{H}^{(i)}_2 \Big) 
       \otimes \mathcal{B}\Big(\bigotimes_{i\in \mathscr{S}' \smallsetminus \mathscr{S}} \mathscr{H}^{(i)}_2 \Big),
  ~~
  \mathcal{P}_{_{\hs{-2pt}\mathscr{S}}}(\pi):  \mathscr{H} \rightarrow \mathscr{H}.
\end{eqnarray}

\noindent%
The notion of `permutationally-symmetric states' was introduced by T\'oth and G\"uhne for a compound system of two $d$-state bosons \cite{Toth&Gruehne:PRL:2009}. Below, we give the definition for compound systems of $n$ qubits.

\begin{definition}[Permutationally-symmetric qubit string]\label{def:permutational_symmetric_Qubit-String}
  A qubit string $\rho \in \mathcal{B}\big(\mathscr{H}_2^{\otimes n}\big)$ that satisfies 
    	$\rho =  \mathcal{P}(\pi) \rho  =  \rho \mathcal{P}(\pi)^\dagger$
   for all $\pi \in S_{n}$ is permutationally-symmetric.
\end{definition}

\noindent%
Definition \ref{def:permutational_symmetric_Qubit-String} implies that a pure state $\ket{\Psi} \in \mathscr{H}_2^{\otimes n}$ is permutationally-symmetric if and only if $\forall\pi \in S_{n}:  \ket{\Psi} = \mathcal{P}(\pi)  \ket{\Psi}$.

A qubit string that is not permutationally-symmetric could, however, be permutationally-symmetric in a subspace. Consequently, we define permutation symmetry over a set of qubits below.

\begin{definition}[Permutation symmetry over a set $\mathscr{S}$ of qubits]\label{def:permutational_symmetric_Qubit-String_for_Set}
  Let  $\mathscr{S} \subseteq \{1,2,\ldots,n\}$. 
  An $n$-qubit string $\rho \in \mathcal{B}\big(\mathscr{H}_2^{\otimes n}\big)$ that satisfies 
    	$\rho =  \mathcal{P}(\pi_\mathscr{S}) \rho  =  \rho \mathcal{P}(\pi_\mathscr{S})^\dagger$
   for all $\pi_\mathscr{S} \in S_{\mathscr{S}}$, is permutationally-symmetric over the set $\mathscr{S}$.
\end{definition}

\subsection{Basis states for permutationally-symmetric qubit strings}

\noindent%
The Hilbert space of permutationally-symmetric qubit strings of length $n$ forms a subspace of $\mathscr{H}_2^{\otimes n}$. In Definition \ref{def:permutationally_symmetric_state} below, we construct $n+1$ orthonormal permutationally-symmetric states. 
These states are identical to the `symmetric states' \cite{Browne&Plenio:PhysRevA.66.2002,Stockton&al:PhysRevA.67.2003}. 
In contrast, we sum over all permutations in Definition \ref{def:permutationally_symmetric_state} in order to exploit the group properties of $S_n$ whereas Stockton et al.\ only consider distinct permutations \cite{Stockton&al:PhysRevA.67.2003}. 

\begin{definition}[Permutationally-symmetric state ($\pi$-state)]\label{def:permutationally_symmetric_state}
For the set $\mathcal{U}=\{u_1,u_2,\ldots,u_n\} \subset \mathbbm{Z}_+$ of qubit positions, with $u_n > \cdots > u_2 > u_1$, the permutationally-symmetric $n$-qubit state is
\begin{eqnarray} \label{eq:orderInvaiantState_Raw}
	\PiStateFull{\nu}{\mathcal{U}} &:= \frac{1}{\sqrt{n!\, \nu!\, (n-\nu)!}} \sum_{\pi \in S_n} \mathcal{P}(\pi) ~
				\big( \ket{0}_{u_n}\cdots \ket{0}_{u_{\nu +1}} \, \ket{1}_{u_{\nu}} \cdots \ket{1}_{u_1} \big) 
	~ \in \bigotimes_{i\in \mathcal{U}}\mathscr{H}^{(i)}_2,
\end{eqnarray}
for $\ket{\beta}_u$ the state of the qubit with position $u \in \mathbbm{Z}_+$.
If $\mathcal{U} = \{1,2,\ldots,n\}$, we write $\PiStateFull{\nu}{n}$.
\end{definition}

\noindent%
There exists a bijective linear correspondence that maps every state $\ket{\beta_n\ldots \beta_1} \in  \bigotimes_{i\in \mathcal{U}} \mathscr{H}^{(i)}_2$ to $\ket{\beta_n\ldots \beta_1} \in  \bigotimes_{i\in \{1,\ldots,n\}} \mathscr{H}^{(i)}_2$. In most cases, this bijection allows us, without loss of generality, to work with the states $\PiStateFull{\nu}{n}$, which comprises qubits with `default' positions $1,2,\ldots,n$.

For a compound system of $n$ spin-$\frac{1}{2}$ systems, i.e.\ qubits, the Dicke states \cite{Dicke.PhysRev.93.99} are defined as the states $\ket{l,m}_\tn{D}$ that are simultaneous eigenstates of both the square of the total spin operator $\widehat{\mathitbf{J}}^2$ and its $z$ component $\widehat{J}_z$ \cite{Mandel&Wolf:Optical_Coherence_and_Quantum_Optics}:
\begin{eqnarray}
	\widehat{\mathitbf{J}}^2 \ket{l,m}_\tn{D} = l(l+1)\ket{l,m}_\tn{D}, 
  \qquad	\widehat{J}_z\ket{l,m}_\tn{D} = m\ket{l,m}_\tn{D}.
\end{eqnarray}
The $n+1$ Dicke states $\ket{\frac{n}{2},m}_\tn{D}$, $m \in \{-\frac{n}{2},-\frac{n}{2}+1, \ldots,\frac{n}{2}-1,\frac{n}{2}\}$, form a special subset of all $2^n$ Dicke states in the sense that they are the only states that are symmetric under arbitrary permutations of the qubits \cite{Thiel&al:PRL:2007}. Interpreting  a spin-$\frac{1}{2}$ system with spin in the $+z$ direction as the logical state $\ket{1}$, we have 
\begin{eqnarray}
   \ket{\case{n}{2},m}_\tn{D} \equiv \PiStateFull{\case{n}{2}+m}{n}, 
   \qquad m \in \{-\case{n}{2},-\case{n}{2}+1, \ldots,\case{n}{2}-1,\case{n}{2}\}.
\end{eqnarray}

\begin{remark}\label{remark:Permut.Inv.State_unaffected_by_permutation}
	Equation \eqref{eq:orderInvaiantState_Raw} and Remark \ref{eq:decomposition_of_Permutation_operator} imply that
	\begin{eqnarray}\label{eq:invariance_of_orderInvaiantState}
	\forall\, \pi \in S_n:~ \mathcal{P}(\pi) \PiStateFull{\nu}{n} = \PiStateFull{\nu}{n} \, .
	\end{eqnarray}
\end{remark}

\noindent%
Throughout this paper, we use the notation ``$ \verb|lhs| \stackrel{(x)}{=} \verb|rhs|$'' to indicate that the right-hand side \verb|rhs| has been obtained from the left-hand side \verb|lhs| under the use of equation $(x)$. For instance,
\begin{eqnarray} \label{eq:Fock_states_from_basis}
 	{}_{[n]}\langle\nu\PiStateFull{\mu}{m} \stackrel{\eqref{eq:orderInvaiantState_Raw}}{=} \delta_{m,n}\delta_{\mu,\nu}
\end{eqnarray}
follows from equation \eqref{eq:orderInvaiantState_Raw}.
For fixed $n$, the states $\{\PiStateFull{\nu}{n}\}_{\nu=1,\ldots,n}$ form an orthonormal basis of the $(n+1)$-dimensional Hilbert space
$\mathscr{H}_{n+1}$. Furthermore, $\{\PiStateFull{\nu}{n}\bra{\mu}\}_{\nu,\mu=1,\ldots,n}$ span the space $\mathcal{B}\big(\mathscr{H}_{n+1}\big)$ of linear operators on $\mathscr{H}_{n+1}$, i.e.\
\begin{eqnarray}\label{eq:Hilbert_space_spanned_by_Fock_states}
 	\mathscr{H}_{n+1} :=~ \Span\big\{ \PiStateFull{0}{n} ,\PiStateFull{1}{n} ,\ldots,\PiStateFull{n}{n}  \big\}\,, \qquad
        \mathcal{B}\big(\mathscr{H}_{n+1}\big) =\, \Span\big\{ \PiStateFull{\nu}{n}\bra{\mu}\, : \nu,\mu \in \{0,1,\ldots,n\}\big\}.
\end{eqnarray}
Definition \ref{def:permutationally_symmetric_state} and \eqref{eq:Definition_B(H_2^n)} imply that $\PiStateFull{\nu}{n}\bra{\mu} \in \mathcal{B}\big(\mathscr{H}_{2}^{\otimes n}\big)$, for $\nu,\mu \in \{0,1,\ldots,n\}$. Consequently, $\mathcal{B}\big(\mathscr{H}_{n+1}\big)\subseteq \mathcal{B}\big(\mathscr{H}_{2}^{\otimes n}\big)$ as the basis elements of $\mathcal{B}\big(\mathscr{H}_{n+1}\big)$ are elements of $\mathcal{B}\big(\mathscr{H}_{2}^{\otimes n}\big)$.

In Remark \ref{remark:PiState_has_constant_hamming_weight} below, we use the Hamming weight $\omega$, where $\omega(\mathitbf{b})$ of a bit-vector $\mathitbf{b}= (b_n b_{n-1} \cdots b_1) \in \{0,1\}^{n}$ is the number of non-zero elements in $\mathitbf{b}$ \cite{R.W.Hamming:1950,Fundamentals.Error-Correcting.Codes:Cambridge.2003}.
Furthermore, we abbreviate the action of a permutation $\pi \in S_n$ on the bit-string $\mathitbf{b} \in \{0,1\}^n$ as 
$\pi(\mathitbf{b}) = (b_{\pi^{-1}(n)}\, b_{\pi^{-1}(n-1)}\cdots b_{\pi^{-1}(1)})$.

\begin{remark} \label{remark:PiState_has_constant_hamming_weight}
  The state $ \PiStateFull{\nu}{n}$ is an equally-weighted superposition of all possible qubit strings of length $n$ and Hamming weight $\nu$:
  \begin{eqnarray}\label{eq:Representation_of_Pi_state_in_computational_basis}
		\PiStateFull{\nu}{n} = \binom{n}{\nu}^{-\frac{1}{2}} \sum_{\mathitbf{a} \in \mathcal{W}^{(n)}_\nu} \ket{\mathitbf{a}},
  \end{eqnarray}
  for $\mathcal{W}^{(n)}_\nu := \{ \mathitbf{a} \in \{0,1\}^n: ~\omega(\mathitbf{a}) = \nu \}$ the set of all $n$-bit strings with Hamming weight $\nu$.
\end{remark}

\noindent%
Theorem \ref{theorem:support_only_permut-inv.states} below states that a mixed state $\rho \in \mathcal{B}\big(\mathscr{H}_2^{\otimes n}\big)$ is permutationally-symmetric if and only if it is a linear combination of $\PiStateFull{k}{n}\bra{m}$, $k,m \in \{0,1,\ldots,n\}$. 

\begin{theorem}\label{theorem:support_only_permut-inv.states}
	For all $\rho \in \mathcal{B}\big(\mathscr{H}_2^{\otimes n}\big)$, the following statement holds. 
	\begin{eqnarray}
		\rho \in  \mathcal{B}\big(\mathscr{H}_{n+1}\big) 
	\quad \Leftrightarrow  \quad  
		\forall\, \pi \in S_n: ~ \rho=  \mathcal{P}(\pi) \rho = \rho\, \mathcal{P}(\pi)^\dagger
	\end{eqnarray}
\end{theorem}

\noindent%
Proof of Theorem \ref{theorem:support_only_permut-inv.states}:
\begin{itemize}[leftmargin=27pt] 
 \item[$(\Rightarrow)$] We assume $\rho \in \mathcal{B}\big(\mathscr{H}_{n+1}\big)$. 
			Equation \eqref{eq:Hilbert_space_spanned_by_Fock_states} implies that $\rho$ 
			is a linear combination of permutationally-symmetric states. 
			Therefore, $\rho$ is permutationally-symmetric.
			
 \item[$(\Leftarrow)$] 	We assume that $\rho \in \mathcal{B}\big(\mathscr{H}_2^{\otimes n}\big)$ is permutationally-symmetric, i.e.\ 
			$\forall\, \pi \in S_n: \rho = \mathcal{P}(\pi)\rho = \rho \mathcal{P}(\pi)^\dagger$. 
			Hence, 
			\begin{eqnarray}\label{eq:Proof_theorem:support_only_permut-inv.states-Eq.1}
			 	\frac{1}{(n!)^2} \sum_{\pi, \pi' \in S_n} \mathcal{P}(\pi)\, \rho\, \mathcal{P}(\pi')^\dagger = \rho.
			\end{eqnarray}
			Furthermore, Definition \ref{def:permutationally_symmetric_state} implies that 
 			$\forall \mathitbf{b} \in \{0,1\}^n: \sum_{\pi \in S_n} \mathcal{P}(\pi) \ket{\mathitbf{b}} \in \mathscr{H}_{n+1}$, i.e.
			\begin{eqnarray} \label{eq:Proof_theorem:support_only_permut-inv.states-Eq.2}
			 	& \forall \mathitbf{b},\mathitbf{b}' \in \{0,1\}^n: 
				      \sum_{\pi, \pi' \in S_n} \mathcal{P}(\pi) \ket{\mathitbf{b}}\bra{\mathitbf{b}'} \mathcal{P}(\pi')^\dagger 
				      \in \mathcal{B}\big(\mathscr{H}_{n+1}\big)
			   \\  \label{eq:Proof_theorem:support_only_permut-inv.states-Eq.3}
			   \Rightarrow \quad
				& \forall \rho \in \mathcal{B}\big(\mathscr{H}_2^{\otimes n}\big): ~
				      \sum_{\pi, \pi' \in S_n} \mathcal{P}(\pi)\, \rho\, \mathcal{P}(\pi')^\dagger ~ \in \mathcal{B}\big(\mathscr{H}_{n+1}\big).
			\end{eqnarray}
			Equation \eqref{eq:Proof_theorem:support_only_permut-inv.states-Eq.1} and \eqref{eq:Proof_theorem:support_only_permut-inv.states-Eq.3} together 
			imply that $\rho \in \mathcal{B}\big(\mathscr{H}_{n+1}\big)$.
\end{itemize}
\vs{-15pt} \begin{flushright}  $\square$ \end{flushright}

\begin{corollary}\label{Corollary:Expansion_of_Permut.Inv.State_in_Permut.Inv.Basis}
Any permutationally-symmetric state $\rho$ can be written as 
\begin{eqnarray} \label{eq:Permutation_invariant_state:Basis_expansion}
	\rho = \sum_{\mu,\nu =0}^n \alpha_{\mu,\nu}~ \PiStateFull{\mu}{n}\bra{\nu}
\end{eqnarray}
using the basis states of \eqref{eq:orderInvaiantState_Raw}.
\end{corollary}

\noindent%
Equation \eqref{eq:Permutation_invariant_state:Basis_expansion} provides us with a convenient, compact notation for permutationally-symmetric states. Furthermore, a permutationally-symmetric string of $n$ qubits represents a qudit of dimension $n+1$.

\subsection{Factoring single qubits from a permutationally-symmetric qubit string}

\noindent%
In this section, we show that any permutationally-symmetric state $\PiStateFull{\nu}{n} \in \mathscr{H}_{n +1}$ can be represented as a tensor product of a single qubit and a permutationally-symmetric state $\PiStateFull{\nu'}{n-1} \in \mathscr{H}_{n}$. This is because the tensor product of $\mathscr{H}_{n}$ and a single-qubit Hilbert space 
$\mathscr{H}_{2} =  \Span \{\PiStateFull{0}{1},\PiStateFull{1}{1} \} $ 
can be written as
\begin{eqnarray}
 	 \mathscr{H}_{n}\otimes \mathscr{H}_2 =  \mathscr{H}_{n-1} \oplus  \mathscr{H}_{n+1} \supset \mathscr{H}_{n+1}\, .
\end{eqnarray}

\begin{theorem}\label{theorem:general_split_of_permut.-inv._state}
For $\mathcal{U},\mathcal{V}$ a partition of $\{1,2,\ldots,n\}$, 
\begin{eqnarray}\label{theorem-Eq1:general_split_of_permut.-inv._state}
 & \PiStateFull{\nu}{n}  = \sum_{\mu = 0}^{|\mathcal{V}|} \Xi_{_{|\mathcal{V}|},n; \mu,\nu} ~ \PiStateFull{\nu-\mu}{\mathcal{U}}\otimes \PiStateFull{\mu}{\mathcal{V}}\,,
 \\ \label{theorem-Eq2:general_split_of_permut.-inv._state} 
 & \Xi_{k,n; \mu,\nu} = \binom{n}{\nu}^{-\frac{1}{2}} \binom{n-k}{\nu-\mu}^{\frac{1}{2}} \binom{k}{\mu}^{\frac{1}{2}} \Theta(n-k-\nu+\mu)\Theta(\nu-\mu)
 \\
 & \Theta(x) = \cases{1, & $x \geq 0$ \\ 0, & $x < 0$\,.}
\end{eqnarray}
\end{theorem}

\noindent%
A similar statement to Theorem \ref{corollary:split_of_single_Qubit} was suggested by Gisin and Bechmann-Pasquinucci \cite{Gisin:PLA:19981}, however without a formal proof. Below, we give a rigorous proof based on permutationally-symmetric states. \\

\noindent%
Proof of Theorem \ref{theorem:general_split_of_permut.-inv._state}:\\
First, we note that for any pure $n$-qubit state 
\begin{eqnarray}\label{eq:theorem:general_split_of_permut.-inv._state_Eq.1}
  \sum_{\mathitbf{a} \in \{0,1\}^n} \alpha_\mathitbf{a} \ket{\mathitbf{a}} 
	= \sum_{\kappa = 0}^n \sum_{\mathitbf{a} \in \mathcal{W}^{(n)}_\kappa} \alpha_\mathitbf{a} \ket{\mathitbf{a}}.
\end{eqnarray}
The state $\sum_{\mathitbf{a} \in \mathcal{W}^{(n)}_\nu} \ket{\mathitbf{a}}$  is an equally-weighted superposition of all possible $n$-qubit strings with exactly $\nu$ qubits in the logical state $1$. Partitioning these qubits into two sets $\mathcal{U}$ and $\mathcal{V}$ yields
\begin{eqnarray}\label{eq:theorem:general_split_of_permut.-inv._state_Eq.2}
  \sum_{\mathitbf{a} \in \mathcal{W}^{(n)}_\nu} \ket{\mathitbf{a}} 
	= \sum_{{\mathitbf{b} \in \{0,1\}^{|\mathcal{U}|}} \atop {\mathitbf{b}' \in \{0,1\}^{|\mathcal{V}|}} } 
		\delta_{\omega(\mathitbf{b}) + \omega(\mathitbf{b'}), \nu}~ \ket{\mathitbf{b}}_\mathcal{U}\otimes \ket{\mathitbf{b'}}_\mathcal{V},
\end{eqnarray}
for $ \ket{\mathitbf{b}}_\mathcal{U}$ the collective state of the qubits in the set $\mathcal{U}$. 
Then, \eqref{eq:theorem:general_split_of_permut.-inv._state_Eq.1} implies
\begin{eqnarray}\label{eq:theorem:general_split_of_permut.-inv._state_Eq.3}
\eqalign{
  \sum_{\mathitbf{a} \in \mathcal{W}^{(n)}_\nu} \ket{\mathitbf{a}} 
	& \stackrel{\eqref{eq:theorem:general_split_of_permut.-inv._state_Eq.1}}{=} 
		\sum_{\kappa = 0}^{|\mathcal{U}|} \sum_{\mu = 0}^{|\mathcal{V}|}  \delta_{\kappa + \mu, \nu}
		\sum_{\mathitbf{b} \in \mathcal{W}^{(|\mathcal{U}|)}_\kappa} \sum_{\mathitbf{b}' \in \mathcal{W}^{(|\mathcal{V}|)}_\mu}
		\ket{\mathitbf{b}}_\mathcal{U}\otimes \ket{\mathitbf{b'}}_\mathcal{V}										\\
	& \stackrel{\eqref{eq:Representation_of_Pi_state_in_computational_basis}}{=} 
		\sum_{\kappa = 0}^{|\mathcal{U}|} \sum_{\mu = 0}^{|\mathcal{V}|} \delta_{\kappa + \mu, \nu}
		\binom{|\mathcal{U}|}{\kappa}^{\frac{1}{2}} \binom{|\mathcal{V}|}{\mu}^{\frac{1}{2}} 
		\PiStateFull{\kappa}{\mathcal{U}} \otimes \PiStateFull{\mu}{\mathcal{V}}									\\
	& ~=	\sum_{\kappa = -\infty }^{\infty} \sum_{\mu = 0}^{|\mathcal{V}|} \Theta(\kappa) \Theta(|\mathcal{U}| - \kappa)	 \, \delta_{\kappa + \mu, \nu}
		\binom{|\mathcal{U}|}{\kappa}^{\frac{1}{2}} \binom{|\mathcal{V}|}{\mu}^{\frac{1}{2}} 
		\PiStateFull{\kappa}{\mathcal{U}} \otimes \PiStateFull{\mu}{\mathcal{V}}									\\
	& ~=	\sum_{\mu = 0}^{|\mathcal{V}|}   \Theta\big(\nu-\mu\big) \Theta\big(|\mathcal{U}| - (\nu-\mu)\big)	
		\binom{|\mathcal{U}|}{\nu-\mu}^{\frac{1}{2}} \binom{|\mathcal{V}|}{\mu}^{\frac{1}{2}} 
		\PiStateFull{\nu-\mu}{\mathcal{U}} \otimes \PiStateFull{\mu}{\mathcal{V}}\,.
}
\end{eqnarray}
Since $|\mathcal{U}| = n - |\mathcal{V}|$, one obtains \eqref{theorem-Eq1:general_split_of_permut.-inv._state} from  \eqref{eq:theorem:general_split_of_permut.-inv._state_Eq.3} by multiplying with $\binom{n}{\nu}^{-1/2}$ and using Remark \ref{remark:PiState_has_constant_hamming_weight}.
\vs{-5pt} \begin{flushright}  $\square$ \end{flushright}

\noindent%
As an example for Theorem \ref{theorem:general_split_of_permut.-inv._state}, let us consider the case where we wish to apply some single-qubit unitary transformation $U_1$ to the first qubit of the permutationally-symmetric three-qubit state $\PiStateFull{1}{3}$. Then, for $\mathcal{U} = \{1\}$ and $\mathcal{V}= \{2,3\}$, Theorem \ref{theorem:general_split_of_permut.-inv._state} implies
\begin{eqnarray}
   (U_1 \otimes \mathbbm{1}_{\{2,3\}})\, \PiStateFull{1}{3} 
	= (U_1 \otimes \mathbbm{1}_{\{2,3\}})\, \bigg( 		\sqrt{\frac{1}{3}}\PiStateFull{1}{\{1\}}\PiStateFull{0}{\{2,3\}} 
							+ 	\sqrt{\frac{2}{3}}\PiStateFull{0}{\{1\}}\PiStateFull{1}{\{2,3\}} \bigg),
\end{eqnarray}
for $\mathbbm{1}_{\{2,3\}}$ the identity operation on the second and third qubits. The permutationally-symmetric one-qubit state $\PiStateFull{1}{\{1\}}$ has the following two properties: (1) it consists of the single qubit at the first position and (2) it contains one qubit in the logical state $\ket{1}$. Consequently, $\PiStateFull{1}{\{1\}}$ is identical to state $\ket{1}_{_1}$. The identity $\PiStateFull{\beta}{\{1\}} = \ket{\beta}_{_1}$, for $\beta = 0,1 $, also follows directly from Definition \ref{def:permutationally_symmetric_state}. Thus,
\begin{eqnarray}
   (U_1 \otimes \mathbbm{1}_{\{2,3\}})\, \PiStateFull{1}{3} 
	=  \sqrt{\frac{1}{3}} \big( U_1 \ket{1}_{_1} \big) \PiStateFull{0}{\{2,3\}} + \sqrt{\frac{2}{3}}\big( U_1 \ket{0}_{_1} \big) \PiStateFull{1}{\{2,3\}}\,.
\end{eqnarray}
Our example shows that, using Theorem \ref{theorem:general_split_of_permut.-inv._state}, one can separate qubits from a permutationally-symmetric state without expanding the state into the full computational basis $\PiStateFull{1}{3} = \frac{1}{\sqrt{3}}(\ket{100} + \ket{010} + \ket{001})$. 
This compression is significant for many-qubit states such as $\ket{\Psi} = \sum_{\nu=0}^n \psi_\nu \PiStateFull{\nu}{n}$, $n \gg 1$. In the computational basis, $\PiStateFull{\nu}{n}$ is a superposition of $\binom{n}{\nu}$ logical states and, hence, $\ket{\Psi}$ is a superposition of $\sum_{\nu=0}^n \binom{n}{\nu} = 2^n$ different basis states. If we use the computational basis to apply the single-qubit unitary operator $U_1$ to $\ket{\Psi}$, we work with $2^n$ terms, for $n$ the length of the qubit-string $\ket{\Psi}$. In contrast, making use of the permutation symmetry and Theorem \ref{theorem:general_split_of_permut.-inv._state}, one can work with $2n$ basis states.

Separating a single qubit from a permutationally-symmetric state is important for adaptive-measurement schemes \cite{PhysRevLett.85.5098,PhysRevA.63.053804,Berry-Wiseman:PhysRevA:2009,QLearning:hentschel:PRL:2010,QLearning:hentschel:ITNG:2010}. Therefore, we state the result of Theorem \ref{theorem:general_split_of_permut.-inv._state} for $\mathcal{V} = \{u\} \subset \mathbbm{Z}_+$ as Corollary \ref{corollary:split_of_single_Qubit} below.

\begin{corollary}\label{corollary:split_of_single_Qubit}
Given $\mathcal{U} \subset \mathbbm{Z}_+$, $|\mathcal{U}| =n$, 
the permutationally-symmetric state $\PiStateFull{\nu}{\mathcal{U}}$ can be written in the basis of $\mathscr{H}^{(\mathcal{U}')}_{n}\otimes\mathscr{H}^{(u)}_{2}$, for $\mathscr{H}^{(u)}_{2}$ the single-qubit Hilbert space of the qubit with position $u \in \mathcal{U}$ and $\mathscr{H}^{(\mathcal{U}')}_{n}$ the permutationally-symmetric space of the other $(n-1)$ qubits with positions in $\mathcal{U}' = \mathcal{U} \smallsetminus \{u\}$, as
\begin{eqnarray} \label{eq:gen_split_up_Fock_state} 
\PiStateFull{\nu}{\mathcal{U}}  =		
		\sqrt{\frac{n-\nu}{n}}~ \PiStateFull{\nu}{\mathcal{U}'}\otimes  \ket{0}_{u}
	+ 	\sqrt{\frac{\nu}{n}}~ \PiStateFull{\nu-1}{\mathcal{U}'}\otimes \ket{1}_{u}\,.
\end{eqnarray}
\end{corollary}

\begin{lemma}\label{Lemma:Separated_qubit_can_be_any_qubit}
Given a permutationally-symmetric state $\PiStateFull{\nu}{n} \in \mathscr{H}_{n+1}$, and let $\mathscr{H}^{(i)}_{2}$ be the single-qubit Hilbert space of the $i^\tn{th}$ qubit, with $i \in \{1,2,\ldots,n\}$. Without loss of generality, we can assume that the position of the $i^\tn{th}$ qubit is at the end of the qubit string $\PiStateFull{\nu}{n}$.
\end{lemma}

\noindent%
Proof of Lemma \ref{Lemma:Separated_qubit_can_be_any_qubit}:\\
Let $\widetilde{\pi}\in S_n$ be the permutation that interchanges the $i^\tn{th}$ with the $n^\tn{th}$ qubit. In $\Pop(\widetilde{\pi})\PiStateFull{\nu}{n}$ the formerly  $i^\tn{th}$ qubit now occupies the $n^\tn{th}$ (last) position. However, Remark \ref{remark:Permut.Inv.State_unaffected_by_permutation} states that $\Pop(\widetilde{\pi})\PiStateFull{\nu}{n} = \PiStateFull{\nu}{n}$. Therefore, changing the position of the $i^\tn{th}$ qubit in the string does not affect the state $\PiStateFull{\nu}{n}$. 
\vs{-5pt} \begin{flushright}  $\square$ \end{flushright}

\noindent%
As an example for Lemma \ref{Lemma:Separated_qubit_can_be_any_qubit}, consider measuring qubit 2 of the permutationally-symmetric state $\PiStateFull{1}{3} = \frac{1}{\sqrt{3}}(\ket{100} + \ket{010} + \ket{001}) \in \mathscr{H}_2^{(3)}\otimes \mathscr{H}_2^{(2)}\otimes \mathscr{H}_2^{(1)}$. Lemma 5 implies that the position labels of the qubits in the string $\PiStateFull{1}{3}$ can be arbitrarily chosen. Specifically, we can simply label the last qubit as `qubit 2', i.e.\ we write $\PiStateFull{1}{3}  \in \mathscr{H}_2^{(3)}\otimes \mathscr{H}_2^{(1)}\otimes \mathscr{H}_2^{(2)}$, and then just measure the last qubit of the string.

\begin{definition}[Qubit substring $\tr_{\bar{\varsigma}}(\rho)$ over subset $\varsigma \subseteq \{1,2,\ldots,n\}$]\label{def:qubit_substring}
Let $\rho \in \mathcal{B}(\mathscr{H}_{n+1})$ and $\bar{\varsigma} = \{1,2,\ldots,n\} \smallsetminus \varsigma$. 
The qubit substring over the subset $\varsigma$ is obtained by taking the partial trace over the qubits in $\bar{\varsigma}$:
\begin{eqnarray}
  \tr_{\bar{\varsigma}}(\rho)\, , \qquad
  \tr_{\bar{\varsigma}}:~ \mathcal{B}\Big(\mathscr{H}_2^{\,\otimes n}\Big) \rightarrow \mathcal{B}\Big(\mathscr{H}_2^{\,\otimes (n - |\bar{\varsigma}|)}\Big)
												   = \mathcal{B}\Big(\mathscr{H}_2^{\,\otimes |\varsigma|}\Big)\,.
\end{eqnarray}
Here, $|\bar{\varsigma}|$ denotes the cardinality of the set $\bar{\varsigma}$ 
and $\tr_{\bar{\varsigma}}$ the partial trace over the qubits in $\bar{\varsigma}$.
\end{definition}

\section{Effect of operations on a permutationally-symmetric qubit string}

\noindent%
In this section, we first study the case whereby global permutation symmetry of a permutationally-symmetric qubit string is broken by a quantum process $\mathcal{E}$. 
We show that the remaining qubits after applying $\mathcal{E}$ are in a permutationally-symmetric state.  
Then, we focus on sequential single-qubit measurements and qubit loss for permutationally-symmetric states.

\subsection{Effects of a quantum channel on a permutationally-symmetric qubit string\label{subsec:Effect_General_Channel}}

\noindent%
A general quantum process affecting a quantum state 
$\rho \in \mathcal{B}(\mathscr{H}_2^{\otimes n})$ is conveniently described as a completely-positive, trace-preserving (CPTP) map 
$\mathcal{E}: \mathcal{B}(\mathscr{H}_2^{\otimes n}) \rightarrow \mathcal{B}(\mathscr{H}')$.
The Choi-Jamio\l{}kowski Theorem implies that any CPTP map $\mathcal{E}: \mathbbm{C}^{n\times n} \rightarrow \mathbbm{C}^{m\times m}$ takes the form \cite{Choi1:LinAlgebra:1975,Paulsen:Operator_Algebras:2003}
\begin{eqnarray}
   \mathcal{E}(\rho) = \sum_{k} E_k \rho E_k^\dagger\, , \qquad \sum_{k}  E_k^\dagger E_k = \mathbbm{1}\,.
\end{eqnarray}
The matrices $\{E_k\} \in \mathbbm{C}^{m\times n}$ are the Kraus representation of $\mathcal{E}$. For convenience of notation, we extend the mapping $\mathcal{E}: \mathcal{B}(\mathscr{H}_2^{\otimes n}) \rightarrow \mathcal{B}(\mathscr{H}')$ to any space 
$\mathcal{B}\big(\bigotimes_{i\in \mathscr{S}}\mathscr{H}^{(i)}_2\big)$, $\{1,\ldots,n\} \subseteq\mathscr{S}$, by assuming that $\mathcal{E}$ acts trivially on the subspace $\mathcal{B}\big(\bigotimes_{i\in \mathscr{S}\smallsetminus \{1,\ldots,n\}}\mathscr{H}^{(i)}_2\big)$.

\begin{theorem}\label{theorem:unaffected_qubit-string_permut-invariant}
	Given a permutationally-symmetric state $\rho \in \mathcal{B}\big(\bigotimes_{i\in \mathcal{U}}\mathscr{H}^{(i)}_2\big)$, $ \mathcal{U} \subset \mathbbm{Z}_+$, and a CPTP map $\mathcal{E}$ that acts on the qubits in $\mathscr{S}  \subset \mathcal{U}$, the state $\mathcal{E}(\rho)$ is permutationally-symmetric over the set $\mathcal{U} \smallsetminus \mathscr{S}$.
\end{theorem}

\noindent%
Proof of Theorem \ref{theorem:unaffected_qubit-string_permut-invariant}:\\
The sets $\bar{\mathscr{S}} := \mathcal{U} \smallsetminus \mathscr{S}$ and $\mathscr{S}$ are disjoint subsets of $\mathcal{U}$. 
Let $\{E_k\}$ be the Kraus representation of $\mathcal{E}$. As $\mathcal{E}$ acts trivially on the qubits in $\bar{\mathscr{S}}$, we can write 
$E_k =  E_{\mathscr{S},k}\otimes \mathbbm{1}_{\bar{\mathscr{S}}}$, where $E_{\mathscr{S},k}$ acts solely on $\mathcal{B}\big(\bigotimes_{i\in \mathscr{S}}\mathscr{H}^{(i)}_2\big)$ and $\mathbbm{1}_{\bar{\mathscr{S}}}$ is the identity on 
$\mathcal{B}\big(\bigotimes_{i\in \bar{\mathscr{S}}}\mathscr{H}^{(i)}_2\big)$. Then, for any permutation $\pi_{\bar{\mathscr{S}}} \in S_{|\bar{\mathscr{S}}|}$, we have 
\begin{eqnarray}\fl\nonumber
  \qquad &
 \mathcal{P}(\pi_{\bar{\mathscr{S}}}) \, \mathcal{E}(\rho) \, \mathcal{P}(\pi_{\bar{\mathscr{S}}})^\dagger
 \\\fl &\quad
   = \big( \mathbbm{1}_{\mathscr{S}} \otimes \mathcal{P}(\pi_{\bar{\mathscr{S}}})  \big) 
	\bigg( \sum_k (E_{\mathscr{S},k}\otimes \mathbbm{1}_{\bar{\mathscr{S}}}) \rho (E_{\mathscr{S},k}\otimes \mathbbm{1}_{\bar{\mathscr{S}}})^\dagger \bigg) 
	\big( \mathbbm{1}_{\mathscr{S}} \otimes \mathcal{P}(\pi_{\bar{\mathscr{S}}})  \big)^\dagger
   \\\fl&\quad
   =  \sum_k (E_{\mathscr{S},k}\otimes \mathbbm{1}_{\bar{\mathscr{S}}}) 
		\big( \mathbbm{1}_{\mathscr{S}} \otimes \mathcal{P}(\pi_{\bar{\mathscr{S}}})  \big) 	\rho
		\big( \mathbbm{1}_{\mathscr{S}} \otimes \mathcal{P}(\pi_{\bar{\mathscr{S}}})  \big)^\dagger
	(E_{\mathscr{S},k}\otimes \mathbbm{1}_{\bar{\mathscr{S}}})^\dagger
   \\\fl&\quad
   = \sum_k (E_{\mathscr{S},k}\otimes \mathbbm{1}_{\bar{\mathscr{S}}}) 
	\rho
	(E_{\mathscr{S},k}\otimes \mathbbm{1}_{\bar{\mathscr{S}}})^\dagger
   =  \mathcal{E}(\rho)
\end{eqnarray}
\vs{-15pt} \begin{flushright}  $\square$ \end{flushright}

\noindent%
As the set of all CPTP maps includes partial trace, Theorem \ref{theorem:unaffected_qubit-string_permut-invariant} implies the following corollary.

\begin{corollary}\label{corollary:Pi-State_remains_permut.Inv_under_loss}
A permutationally-symmetric state remains permutationally-symmetric under loss of qubits.
\end{corollary}

\subsection{Sequential single-qubit measurements of a permutationally-symmetric qubit string\label{subsec:Sequential_Single-Qubit_POVMs_and_Loss}}

\noindent%
Here we study the effects of sequential single-qubit measurements and loss on permutationally-symmetric qubit strings. Our analysis provides fundamental insights in the loss robustness of adaptive-measurement schemes that employ permutationally-symmetric qubit strings as input. Specifically we show that qubit loss does not affect adaptive-measurement schemes in the sense that all measurement probabilities are loss-independent. The only consequence of loss is that, trivially, not all qubits can be measured. 

In our analysis, we combine the action of the quantum channel and the measurement into a single positive operator-valued measure (POVM) instead of treating them as separate processes. A POVM is given by its POVM elements $\{\mathcal{K}_\ell\}$, which are positive semi-definite operators such that $\sum_\ell \mathcal{K}_\ell = \mathbbm{1}$. Each POVM element can be decomposed as $\mathcal{K}_\ell = K_\ell^\dagger K_\ell$, where the $\{K_\ell\}$ are the Kraus operators for the measurement. The probability for obtaining the measurement result $\ell$ is
\begin{eqnarray}
 p_\ell = \tr(\mathcal{K}_\ell \rho) = \tr(K_\ell \rho K_\ell^\dagger),
\end{eqnarray}
and the state $\rho_\ell$ after the measurement is
\begin{eqnarray}
 \rho_\ell = \frac{1}{p_\ell} K_\ell \rho K_\ell^\dagger\, .
\end{eqnarray}

In this section, we consider single-qubit POVMs. As we can use the same type of measurement process to measure different qubits within the full string, 
we avoid ambiguity by including a superscript on the Kraus operators. This superscript indicates the qubit position upon which the POVM acts.
For example, the  Kraus operators $\{K^{(3)}_\ell\}$ describe a measurement acting on the third qubit of the full string. In general, the Kraus operators $\{K^{(i)}_\ell\}$ represent a mapping
\begin{eqnarray}
   K^{(i)}_\ell: \mathscr{H}^{(i)}_2 \rightarrow \mathscr{H}^{(i)}_2\,.
\end{eqnarray}
For convenience of notation, we extend the mapping $K^{(i)}_\ell$ to any Hilbert space $\mathscr{H} = \bigotimes_{i\in \mathscr{S}} \mathscr{H}^{(i)}_2$ with $i \in \mathscr{S} \subset \mathbbm{Z}_+$, via 
\begin{eqnarray}\displaystyle
  K^{(i)}_\ell \equiv K^{(i)}_\ell \otimes \mathbbm{1} 
  ~\in~          \mathcal{B}\big(\mathscr{H}^{(i)}_2 \big) 
       \otimes \mathcal{B}\big({\textstyle  \bigotimes_{j\in \mathscr{S} \smallsetminus \{i\}} \mathscr{H}^{(j)}_2} \big)\,,
  \qquad
  K^{(i)}_\ell:  \mathscr{H} \rightarrow \mathscr{H}\,.
\end{eqnarray}

\begin{lemma}\label{lemma:Trace_and_POVM_commpute}
For $\rho \in \mathcal{B}(\mathscr{H}_{n+1})$ a permutationally-symmetric $n$-qubit string and $\{K^{(i)}_\ell\}$ the Kraus operators for a POVM acting on the $i^\tn{th}$ qubit, the partial trace over the $j^\tn{th}$ qubit satisfies
\begin{eqnarray}\label{eq:lemma:Trace_and_POVM_commpute}
     \tr_{\{j\}}  \big( K^{(i)}_\ell \rho K^{(i)\dagger}_\ell \big) 
  =  K^{(i)}_\ell \big(\tr_{\{j\}}   \rho \big)K^{(i)\dagger }_\ell \,.
\end{eqnarray}
\end{lemma}

\noindent%
Proof of Lemma \ref{lemma:Trace_and_POVM_commpute}:\\
If the state is written as $\rho \in \mathcal{B}(\mathscr{H}_2^{\,\otimes n})$, then by using Corollary \ref{Corollary:Expansion_of_Permut.Inv.State_in_Permut.Inv.Basis} and Definition \ref{def:permutational_symmetric_Qubit-String}, it is obvious that the POVM and the partial trace act on non-overlapping subspaces of $\mathcal{B}(\mathscr{H}_2^{\,\otimes n})$. Hence, both operations commute. 
\vs{-5pt} \begin{flushright}  $\square$ \end{flushright}

\begin{remark}\label{theorem:Measurement_Prob._Independent_of_Loss}
   Let $\rho$ be an arbitrary $n$-qubit state and $\{K^{(i)}_\ell\}$ be the Kraus operators for a POVM that acts on the $i^\tn{th}$ qubit of $\rho$. The probability of measurement outcome $\ell$ is independent of how many other qubits have been lost prior to the measurement.
\end{remark}

We now turn to the case where different single-qubit POVMs are sequentially applied to different qubits of a permutationally-symmetric state $\rho$. In order to differentiate between the different POVMs, we introduce a second index $\alpha$ for the Kraus operators. For example, $\{K^{(i)}_{\alpha,\ell}\}$ is the set of Kraus operators for the POVM of type `$\alpha$' acting on the $i^\tn{th}$ qubit of $\rho$. The second index, $\ell$, corresponds to the different measurement outcomes. The same POVM $\alpha$ applied to the $j^\tn{th}$ qubit is denoted by the Kraus operators $\{K^{(j)}_{\alpha,\ell}\}$.

For example, one could measure the $i^\tn{th}$ qubit of an $n$-qubit string in the computational basis $\{\ket{0},\ket{1}\}$ or in the Hadamard basis $\{\ket{0'}, \ket{1'}\}$, with $\ket{0'}=(\ket{0}+\ket{1})/\sqrt{2}$ and $\ket{1'}=(\ket{0}-\ket{1})/\sqrt{2}$. Hence, we are given two sets of Kraus operators, i.e.\ 
$\{K^{(i)}_{1,0} = \ket{0}_i\bra{0}, K^{(i)}_{1,1} = \ket{1}_i\bra{1} \}$ for the measurement in the computational basis and 
$\{K^{(i)}_{2,0} = \ket{0'}_i\bra{0'}, K^{(i)}_{2,1} = \ket{1'}_i\bra{1'} \}$ for the Hadamard basis. 

\begin{theorem}\label{theorem:irrelevant_which_qubit_is_measured}
	Let $\rho$ be a permutationally-symmetric $n$-qubit string that is subject to a time-ordered sequence of $k \in \{0,1,2,\ldots,n\}$ single-qubit measurements with the restriction that no qubit is measured more than once. The individual measurement probabilities are equal to an experiment where the measurements are applied in the same order to the qubits $1,2,\ldots,k$ of $\rho$.
\end{theorem}

\noindent%
Proof of Theorem \ref{theorem:irrelevant_which_qubit_is_measured}: \\
Let the number of measurements performed on $\rho$ be $M \in \{1,2,\ldots,n\}$, with $\{K^{(s_1)}_{1,\ell}\}$ the Kraus operators for the first measurement, $\{K^{(s_2)}_{2,\ell}\}$ the Kraus operators for the second measurement, and so on, up to $\{K^{(s_M)}_{M,\ell}\}$ the Kraus operators for the last measurement. The set of measured qubits is then given by $\mathcal{S} = \{s_1,s_2,\ldots,s_M\} \subseteq \{1,2,\ldots,n\}$. As no qubit is measured more than once, i.e.\ $s_i \neq s_j$ for all 
$i,j \in \{1,\ldots,n\}$ with $i \neq j$. 
The probability for obtaining outcome $\ell_1$ in the first measurement, $\ell_2$ in the second measurement up to $\ell_M$ in the $M^\tn{th}$ measurement is
\begin{eqnarray}\label{theorem:irrelevant_which_qubit_is_measured_eq_1}
   p_\mathcal{S}(\ell_M, \ldots, \ell_{2}, \ell_1) 
 = \tr\Big(  K^{(s_M)}_{M,\ell_M}\cdots K^{(s_1)}_{1,\ell_1} \, \rho \, K^{(s_1)\dagger}_{1,\ell_1} \cdots K^{(s_M)\dagger}_{M,\ell_M} \Big) \,.
\end{eqnarray}
We now prove
\begin{eqnarray}\label{theorem:substatement_1}\fl
   &\tr\Big(  K^{(s_M)}_{M,\ell_M}\cdots K^{(s_1)}_{1,\ell_1} \, \rho \, K^{(s_1)\dagger}_{1,\ell_1} \cdots K^{(s_M)\dagger}_{M,\ell_M} 		\Big)
  = \tr\Big(  K^{(s'_M)}_{M,\ell_M} \cdots K^{(s'_2)}_{2,\ell_2} \,\rho' \, K^{(s'_2)\dagger}_{2,\ell_2} \cdots  K^{(s'_M)\dagger}_{M,\ell_M}      \Big)\,,
\\\fl \label{theorem:substatement_1_definitions}
&\eqalign{
	&\rho' := \tr_{\{n\}}(  K^{(n)}_{1,\ell_1} \,\rho\, K^{(n)\dagger}_{1,\ell_1})\,, 
	\\
	 &(s'_2,\ldots,s'_M) := \cases{
			            	(s_2,s_3,\ldots,s_M)\,,				& $\nexists m \in \{2,\ldots,M\}:  s_m = n$ \\
			            	(s_2,\ldots,s_{m-1},s_1,s_{m+1},\ldots,s_M)\,, 	& $\exists  m \in \{2,\ldots,M\}:  s_m = n$\,.
			         }
}
\end{eqnarray}
We now proceed to proving equation \eqref{theorem:substatement_1}. 

Let $\Xi_{k,m}: \mathscr{H}_2^{\,\otimes n} \rightarrow \mathscr{H}_2^{\,\otimes n}$, $k,m \in \{1,2,\ldots,n\}$ be the swap operation that swaps the qubit at position $k$ in an $n$-qubit string with the one at position $m$. For any $n$-qubit state, one can either measure the $j^\tn{th}$ qubit directly or swap qubits $k$ and $j$, measure qubit $k$, and finally swap qubits $k$ and $j$ back:
\begin{eqnarray}\label{proof:Lemma:Loss_and_POVM_commute_eq0}
    K^{(j)}_\ell  = \Xi^\dagger_{k,j} \, K^{(k)}_\ell \, \Xi_{k,j} \,.
\end{eqnarray}
The swap operation, simply permuting two qubits, does not change a permutationally-symmetric state; i.e.
\begin{eqnarray}\label{eq:permut.-inv._states_are_SWAP-invariant}
 	\rho  = \Xi_{k,j} \, \rho\, \Xi^\dagger_{k,j}\,.
\end{eqnarray}
Furthermore, $\Xi_{k,m}$ commutes with all operations that do not involve qubit $k$ or qubit $m$. Specifically,
\begin{eqnarray}
   \forall i \in \{1,2\ldots,n\} \smallsetminus \{k,m\}: ~ 
	\big[ \Xi_{k,m}\,,\, K^{(i)}_{\alpha,\ell} \big] = \big[ \Xi_{k,m}\,,\,  K^{(i)\dagger}_{\alpha,\ell} \big] = 0\,.
\end{eqnarray}
Using $\Xi_{k,m} = \Xi^\dagger_{k,m} = \Xi^{-1}_{k,m}$, and the cyclicity of the trace, we can write 
\begin{eqnarray}\fl\label{theorem:irrelevant_which_qubit_is_measured_eq2}
&\eqalign{
 p_\mathcal{S}(\ell_M, \ldots, \ell_1)  
 \stackrel{\eqref{proof:Lemma:Loss_and_POVM_commute_eq0}}{=}
   \tr\Big(
	&\Xi_{s_1,n} ~ K^{(s_M)}_{M,\ell_M} \cdots K^{(s_2)}_{2,\ell_2} 
	~ \Xi_{s_1,n}  K^{(n)}_{1,\ell_1} \Xi_{s_1,n}  ~\rho~ 
	\\& 
	\times \Xi_{s_1,n} K^{(n)\dagger}_{1,\ell_1} \Xi_{s_1,1}~K^{(s_2)\dagger}_{2,\ell_2} \cdots  K^{(s_M)\dagger}_{M,\ell_M} ~ \Xi_{s_1,n} 
      \Big)
}
 \\\fl&\qquad
   \stackrel{\eqref{eq:permut.-inv._states_are_SWAP-invariant}}{=}
   \tr\Big( 
	\Xi_{s_1,n} ~ K^{(s_M)}_{M,\ell_M} \cdots K^{(s_2)}_{2,\ell_2} 
	~ \Xi_{s_1,n}  K^{(n)}_{1,\ell_1} \,\rho\, K^{(n)\dagger}_{1,\ell_1} \Xi_{s_1,n}~
	K^{(s_2)\dagger}_{2,\ell_2} \cdots  K^{(s_M)\dagger}_{M,\ell_M} ~ \Xi_{s_1,n} 
      \Big)\,.
\end{eqnarray}
There are two possible cases.
\begin{itemize}[leftmargin=21pt]
 \item[(a)] 	$n \notin \{s_2,s_3,\ldots,s_{M}\}$, i.e.\ $\nexists m \in \{2,\ldots,M\}\}:  s_m = n$.\\
			Hence, $\Xi_{s_1,n}$ commutes with $K^{(s_{M})}_{{M},\ell_{M}}, \ldots, K^{(s_2)}_{2,\ell_2}$. The fact that $\Xi^2_{k,m} = \mathbbm{1}$ yields
			\begin{eqnarray}\fl
			    \quad p_\mathcal{S}(\ell_M, \ldots, \ell_1) &
			       ~=  \tr\Big( 
					K^{(s_M)}_{M,\ell_M} \cdots K^{(s_2)}_{2,\ell_2} 
					K^{(n)}_{1,\ell_1} \,\rho\, K^{(n)\dagger}_{1,\ell_1} 
					K^{(s_2)\dagger}_{2,\ell_2} \cdots  K^{(s_M)\dagger}_{M,\ell_M}
      				   \Big)
			   \\& \fl\qquad
			        \stackrel{\eqref{eq:lemma:Trace_and_POVM_commpute}}{=}
			 	\tr_{\{1,2,\ldots,n-1\}}\Big( 
					K^{(s_M)}_{M,\ell_M} \cdots K^{(s_2)}_{2,\ell_2} 
					~\tr_{\{n\}}\big(  K^{(n)}_{1,\ell_1} \,\rho\, K^{(n)\dagger}_{1,\ell_1} \big) ~
					K^{(s_2)\dagger}_{2,\ell_2} \cdots  K^{(s_M)\dagger}_{M,\ell_M}
      				    \Big)
			   \\& \fl\qquad
			        \stackrel{\eqref{theorem:substatement_1_definitions}}{=}
				\tr\Big( 
					K^{(s'_M)}_{M,\ell_M} \cdots K^{(s'_2)}_{2,\ell_2} ~\rho' ~ K^{(s'_2)\dagger}_{2,\ell_2} \cdots  K^{(s'_M)\dagger}_{M,\ell_M}
      				    \Big)\,. 
			\end{eqnarray}
			Furthermore, note that $\{s'_2,\ldots,s'_{M}\} = \{s_2,\ldots,s_{M}\} \subseteq \{1,2,\ldots,n-1\}$ because $n \notin \{s_2,s_3,\ldots,s_{M}\}$.
 \item[(b)] 	$n \in \{s_2,s_3,\ldots,s_{M}\}$, i.e.\  $\exists m \in \{2,\ldots,M\}:  s_m = n$.\\
			Consequently, $\forall j \in \{2,3,\ldots,M\} \smallsetminus\{m\}:  [\Xi_{s_1,1}\,,\, K^{(s_j)}_{j,\ell_j}] =0 $. Thus,
			\begin{eqnarray}\fl
			  \eqalign{
			   & p_\mathcal{S}(\ell_M, \ldots, \ell_1) = 
			   \\
			   & \qquad \tr\Big( 
					K^{(s_M)}_{M,\ell_M} \cdots K^{(s_{m+1})}_{m+1,\ell_{m+1}} ~\Xi_{s_1,n} K^{(s_m)}_{m,\ell_m} \Xi_{s_1,n} ~
					K^{(s_{m-1})}_{m-1,\ell_{m-1}} \cdots K^{(n)}_{1,\ell_1} \,\rho\, 
			   \\ \fl & \qquad \quad ~~ 
					\times K^{(n)\dagger}_{1,\ell_1} \cdots  K^{(s_{m-1})\dagger}_{m-1,\ell_{m-1}} 
					~\Xi_{s_1,n} K^{(s_m)\dagger}_{m,\ell_m} \Xi_{s_1,n} ~ K^{(s_{m+1})\dagger}_{m+1,\ell_{m+1}} \cdots K^{(s_M)\dagger}_{M,\ell_M}
      				    \Big)\,.
			  }
			\end{eqnarray}
			As $s_m = n$, we have
			$\Xi_{s_1,n} K^{(s_m)}_{m,\ell_m} \Xi_{s_1,n} \stackrel{\eqref{proof:Lemma:Loss_and_POVM_commute_eq0}}{=} K^{(s_1)}_{m,\ell_m}$, i.e.\
			\begin{eqnarray}\fl
			  &\eqalign{
			   p_\mathcal{S}(\ell_M, \ldots, \ell_1)   
			   &~=
			   \tr\Big( 
					K^{(s_M)}_{M,\ell_M} \cdots K^{(s_{m+1})}_{m+1,\ell_{m+1}} K^{(s_1)}_{m,\ell_m}
					K^{(s_{m-1})}_{m-1,\ell_{m-1}} \cdots K^{(n)}_{1,\ell_1} \,\rho\, 
				\\\fl
				& \qquad\quad
					\times K^{(n)\dagger}_{1,\ell_1} \cdots  K^{(s_{m-1})\dagger}_{m-1,\ell_{m-1}} 
					K^{(s_1)}_{m,\ell_m} K^{(s_{m+1})\dagger}_{m+1,\ell_{m+1}} \cdots K^{(s_M)\dagger}_{M,\ell_M}
      				    \Big)
			  }
			  \\ \fl & \qquad \qquad \qquad
			  \stackrel{\eqref{theorem:substatement_1_definitions}}{=}
				\tr\Big( 
					K^{(s'_M)}_{M,\ell_M} \cdots K^{(s'_2)}_{2,\ell_2} 
					~\rho' ~
					K^{(s'_2)\dagger}_{2,\ell_2} \cdots  K^{(s'_M)\dagger}_{M,\ell_M}
      				    \Big).
			\end{eqnarray}
			Per definition, $\{s_1,s_2,\ldots,s_M\} \subseteq \{1,2,\ldots,n\}$. Consequently,
			$\{s'_2,\ldots,s'_{M}\} = \{s_1,\ldots,s_{m-1},s_{m+1},\ldots,s_M\} \subseteq \{1,2,\ldots,n-1\}$, because $s_m =n$.
\end{itemize}
At this point we have shown 
\begin{eqnarray}\label{theorem:irrelevant_which_qubit_is_measured_eq_3}
   p_\mathcal{S}(\ell_M, \ldots, \ell_1)
   = \tr\Big(  K^{(s'_M)}_{M,\ell_M} \cdots K^{(s'_2)}_{2,\ell_2} \,\rho' \, K^{(s'_2)\dagger}_{2,\ell_2} \cdots  K^{(s'_M)\dagger}_{M,\ell_M}      \Big)\,.
\end{eqnarray}
Equation \eqref{theorem:substatement_1} follows from \eqref{theorem:irrelevant_which_qubit_is_measured_eq_1} and \eqref{theorem:irrelevant_which_qubit_is_measured_eq_3}.

We now complete the proof of Theorem \ref{theorem:irrelevant_which_qubit_is_measured}.
Theorem \ref{theorem:unaffected_qubit-string_permut-invariant} states that 
$\rho' \in \mathcal{B}(\bigotimes_{i\in \{n-1,\ldots,2,1\}} \mathscr{H}^{(i)}_2 ) $ is permutationally-symmetric. Hence, we can recursively apply  \eqref{theorem:substatement_1} to obtain
\begin{eqnarray}\label{theorem:irrelevant_which_qubit_is_measured_eq_4}\fl
\eqalign{
 p_\mathcal{S}(\ell_M, \ldots, \ell_1)  & 
    ~= \tr_{\{n-M+1\}}\Big(  	K^{(n-M+1)}_{M,\ell_M} \cdots 
		\tr_{\{n-1\}}\Big(  K^{(n-1)}_{2,\ell_2} 
   \\ & \quad\qquad\qquad
		\times \tr_{\{n\}}\Big(  K^{(n)}_{1,\ell_1} \,\rho\, K^{(n)\dagger}_{1,\ell_1} \Big) K^{(n-1)\dagger}_{2,\ell_2} \Big)
	  	\cdots  K^{(n-M+1)\dagger}_{M,\ell_M}      \Big)\,.
 \\\fl &\,
    \stackrel{\eqref{eq:lemma:Trace_and_POVM_commpute}}{=}
	\tr\Big(  	K^{(n-M+1)}_{M,\ell_M} \cdots  K^{(n-1)}_{2,\ell_2} K^{(n)}_{1,\ell_1} \,\rho\, 
			K^{(n)\dagger}_{1,\ell_1} K^{(n-1)\dagger}_{2,\ell_2} \cdots  K^{(n-M+1)\dagger}_{M,\ell_M}      \Big)\,.
}
\end{eqnarray}
Equation \eqref{theorem:irrelevant_which_qubit_is_measured_eq_4} states that the measurement probability $p_\mathcal{S}(\ell_M, \ldots, \ell_1)$ for measuring the qubits with positions in $\mathcal{S}$ equals the measurement probability $p_\mathcal{L}(\ell_M, \ldots, \ell_1)$ for measuring the qubit in $\mathcal{L}:= \{n-M+1, \ldots, n-1,n\}$, i.e.\ measuring the last $M$ qubits of the string. Thus, for any two sets $\mathcal{S}, \mathcal{V} \subseteq \{1,2,\ldots,n\}$ with $|\mathcal{S}|=|\mathcal{V}|$ we have
\begin{eqnarray}
   p_\mathcal{S}(\ell_M, \ldots, \ell_1) = p_{ \{n-M+1, \ldots, n-1,n\}}(\ell_M, \ldots, \ell_1) = p_\mathcal{V}(\ell_M, \ldots, \ell_1)\,.
\end{eqnarray}
\vs{-5pt} \begin{flushright}  $\square$ \end{flushright}

\begin{corollary}\label{theorem:Loss_can_be_posponed_to_the_End}
	Let $\rho$ be a permutationally-symmetric $n$-qubit string that is subject to a time-ordered sequence of single-qubit measurements and losses of qubits with restrictions
	\begin{itemize}
	\item	No qubit is measured more than once,
	\item	No qubit that becomes lost has been previously measured.
	\end{itemize}
	Let $k \in \{0,1,2,\ldots,n\}$ be the number of successful single-qubit measurements performed on $\rho$. The individual measurement probabilities are equal to a \emph{lossless} experiment where the measurements are applied in the same order to the qubits $1,2,\ldots,k$ of $\rho$.
\end{corollary}

\noindent%
Proof of Corollary \ref{theorem:Loss_can_be_posponed_to_the_End}: \\
Remark \ref{theorem:Measurement_Prob._Independent_of_Loss} implies that the measurement probabilities are equal to the lossless case. Thus, Theorem \ref{theorem:irrelevant_which_qubit_is_measured} shows that, without altering the measurement probabilities, we can apply the sequence of measurements to the qubits $1,2,\ldots,k$ of $\rho$.
\vs{-5pt} \begin{flushright}  $\square$ \end{flushright}


\section{Implications for adaptive-measurement schemes}

\begin{figure}[t]
    \centering
    \includegraphics[width=0.3\textwidth]{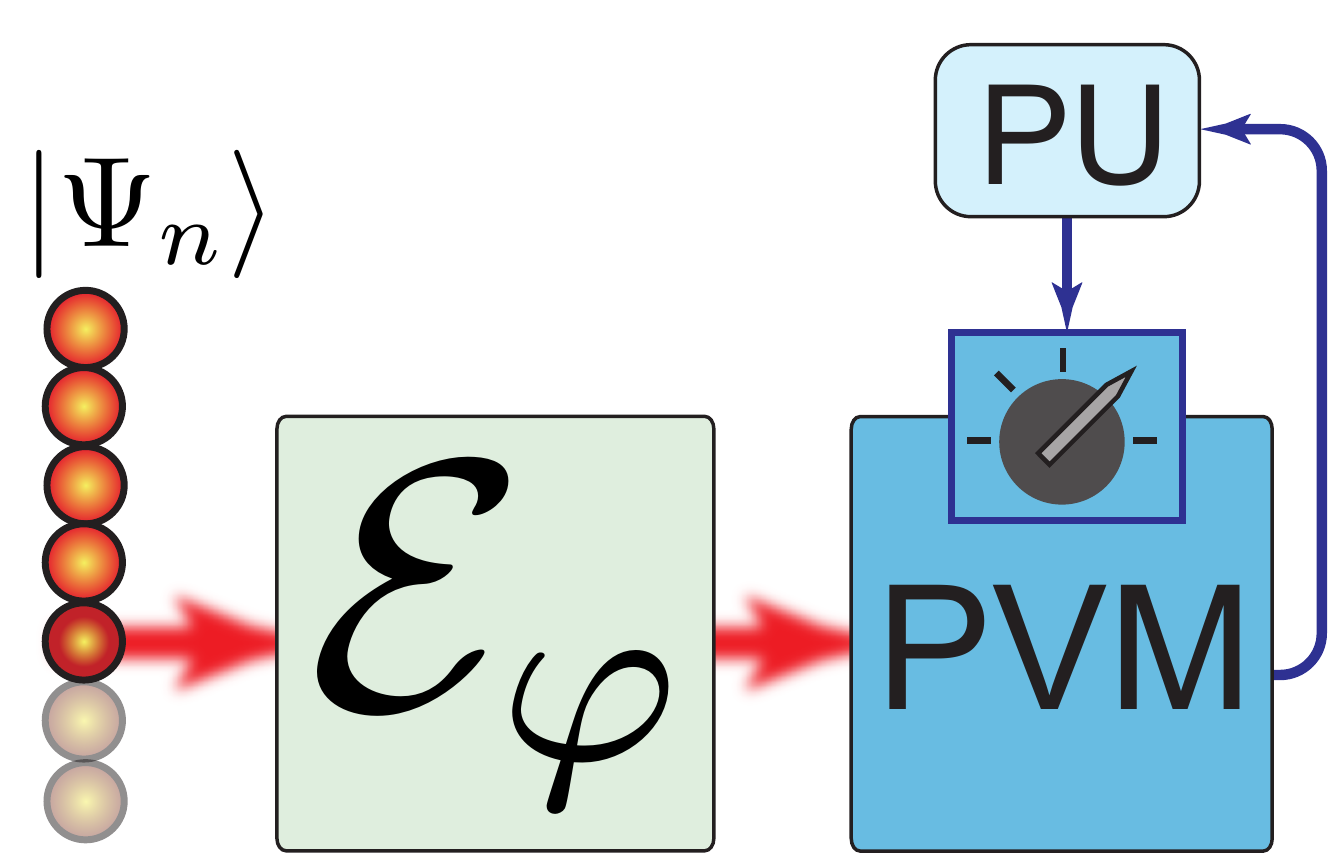}
    \caption{\textbf{Example of an adaptive-measurement scheme.} The task is to estimate the parameter $\varphi$ that governs the transformation by the unitary, single-qubit quantum channel $\mathcal{E}_\varphi$. A permutationally-symmetric $n$-qubit input state $\ket{\Psi_N}$ is stored in a quantum memory, and one qubit at a time is sent through the the channel $\mathcal{E}_\varphi$. The channel output reaches a detector whose operation is described by a projective-valued measure (PVM). Depending on the measurement outcome, a processing unit (PU) adjusts the basis of the PM. Red and blue arrows represent quantum and classical information respectively.
    }
    \label{fig:Adaptive_Feedback_Measurement_Scheme}
\end{figure}

In this section, we state the implications of our results for adaptive-measurement schemes. Adaptive interferometric phase estimation has been studied in depth \cite{PhysRevLett.85.5098,PhysRevA.63.053804,Berry-Wiseman:PhysRevA:2009,QLearning:hentschel:PRL:2010,QLearning:hentschel:ITNG:2010}, and these schemes rely on permutationally-symmetric input states. In fact, it has been shown that permutationally-symmetric states are the optimal inputs \cite{Quantum_phase_estimation_with_lossy_interferometers:PRA:2009}. Furthermore, all schemes so far \cite{PhysRevLett.85.5098,PhysRevA.63.053804,Berry-Wiseman:PhysRevA:2009,QLearning:hentschel:PRL:2010,QLearning:hentschel:ITNG:2010} use projective measurement at the interferometer output, which are experimentally simple to implement. Therefore, we focus on a setting where pure permutationally-symmetric input states and projective-valued measures (PVM) are used to estimate parameters of a unitary single-qubit quantum channel $\mathcal{E}$. Illustration of such estimation schemes is given in Figure \ref{fig:Adaptive_Feedback_Measurement_Scheme}. Our work has two major implications for adaptive-measurement schemes that follow the paradigm of Figure \ref{fig:Adaptive_Feedback_Measurement_Scheme}, such as  \cite{PhysRevLett.85.5098,PhysRevA.63.053804,Berry-Wiseman:PhysRevA:2009,QLearning:hentschel:PRL:2010,QLearning:hentschel:ITNG:2010}. We begin with a remark and then present the implications in subsections \ref{sec:Adaptive_Feedback_Measurement_Schemes_A} and \ref{sec:Adaptive_Feedback_Measurement_Schemes_B}.

As $\mathcal{E}$ is a unitary channel and the measurement is a PVM, the combined action of the channel and the measurement is again a projective-valued single-qubit measurement. We represent this measurement by the two Kraus operators $\{K_0,K_1\}$, corresponding to the binary measurement outcomes $\ell \in \{0,1\}$. As the combined measurement is a PVM, $K_0$ and $K_1$ are projectors. In Remark \ref{Remark:Resulting_State_after_single_qubit_PVM} below, we denote $\ket{0'}, \ket{1'} \in \mathscr{H}_2$ as the combined eigenbasis of $\{K_0,K_1\}$, i.e.\ $K_\ell = \ket{\ell'}\bra{\ell'}$. Remark \ref{Remark:Resulting_State_after_single_qubit_PVM} follows directly from Corollary \ref{corollary:split_of_single_Qubit}.

\begin{remark}\label{Remark:Resulting_State_after_single_qubit_PVM}
   Let 
   \begin{eqnarray}
	    \rho = \sum_{\nu,\mu =0}^n \alpha_{\nu,\mu} ~ \PiStateFull{\nu}{n}\bra{\mu} ~ \in \mathcal{B}(\mathscr{H}_{n+1})
   \end{eqnarray}
   be a permutationally-symmetric $n$-qubit string and $\{K^{(i)}_{0},K^{(i)}_{1}\}$ the Kraus operators for a single-qubit PVM for measuring the $i^\tn{th}$ qubit of $\rho$.
   For the measurement result $\ell \in \{0,1\}$, the state after the measurement is given by  
   \begin{eqnarray}\label{eq:Remark:Resulting_State_after_single_qubit_PVM-Eq1}
      \widetilde{\rho}_\ell = 	\frac{1}{p_\ell}
			     \bigg( \sum_{\nu,\mu =0}^{n-1} \widetilde{\alpha}_{\nu,\mu} \PiStateFull{\nu}{n-1}\bra{\mu} \bigg) 
				\otimes \ket{\ell'}_i\bra{\ell'}
			 \quad  \in 	\mathcal{B}(\mathscr{H}_{n})\otimes\mathcal{B}(\mathscr{H}_{2})\,,	
      \\ \label{eq:Remark:Resulting_State_after_single_qubit_PVM-Eq2}
    \eqalign{
      \widetilde{\alpha}_{\nu,\mu} =  n^{-1}  \big(~ &\, \alpha_{\nu,\mu}     \sqrt{(n-\nu)(n-\mu)} \,\kappa_{\ell,0}\kappa^*_{\ell,0}    
						+  	    \alpha_{\nu,\mu+1}   \sqrt{(n-\nu)(\mu+1)} \,\kappa_{\ell,0}\kappa^*_{\ell,1}		\\
				      		  	& +\, \alpha_{\nu+1,\mu}   \sqrt{(\nu+1)(n-\mu)} \,\kappa_{\ell,1}\kappa^*_{\ell,0}  
						+  	    \alpha_{\nu+1,\mu+1} \sqrt{(\nu+1)(\mu+1)} \,\kappa_{\ell,1}\kappa^*_{\ell,1}		~\big)\,,
    }
   \end{eqnarray}
   with $\kappa_{\ell,b} = {}_i\braket{\ell'}{b}_i$. The measurement probability is $p_\ell = \sum_{\nu=0}^{n-1}  \widetilde{\alpha}_{\nu,\nu}$.
\end{remark}

\noindent
In the following Corollary, we consider the case of a sequence of single-qubit measurements and qubit losses, but with the difference that the measurements are described by PVM's instead of more general POVM's as in Corollary \ref{theorem:Loss_can_be_posponed_to_the_End}. Then, the restriction that ``no qubit that becomes lost has been previously measured'' of Corollary \ref{theorem:Loss_can_be_posponed_to_the_End} can be discarded. 

\begin{corollary}\label{theorem:Loss_can_be_posponed_to_the_End_and_measured_qubits_can_be_lost}
	Let $\rho$ be a permutationally-symmetric $n$-qubit string that is subject to a time-ordered sequence of single-qubit \emph{projective} measurements and losses of qubits with the restriction that no qubit is measured more than once. Let $k \in \{0,1,2,\ldots,n\}$ be the number of successful single-qubit measurements performed on $\rho$. The individual measurement probabilities are equal to a \emph{lossless} experiment where the measurements are applied in the same order to the qubits $1,2,\ldots,k$ of $\rho$.
\end{corollary}

\noindent%
Proof of Corollary \ref{theorem:Loss_can_be_posponed_to_the_End_and_measured_qubits_can_be_lost}: \\
Corollary \ref{theorem:Loss_can_be_posponed_to_the_End_and_measured_qubits_can_be_lost} extends Corollary \ref{theorem:Loss_can_be_posponed_to_the_End} by stating that also previously measured qubits can be lost without altering the subsequent measurement probabilities. Remark \ref{Remark:Resulting_State_after_single_qubit_PVM} shows that the measured qubits are in a product state to the unmeasured qubits. Consequently, the loss of measured qubits has no effect on the outcome of subsequent measurements that are applied to the unmeasured qubits. 
\vs{-5pt} \begin{flushright}  $\square$ \end{flushright}

The exact nature of the loss mechanism is irrelevant as long as the loss process affects only the qubits that are finally lost. To illustrate this, consider the case wherein the second qubit of the three-qubit state $\rho \in \mathcal{B}(\mathscr{H}^{(3)}_2\otimes\mathscr{H}^{(2)}_2\otimes\mathscr{H}^{(1)}_2)$ is lost. As part of the loss process, qubit $2$ undergoes an evolution represented by a CPTP map $\mathcal{C}: \mathcal{B}(\mathscr{H}^{(2)}_2) \rightarrow \mathcal{B}(\mathscr{H}^{(2)}_2)$. Let $\{E_k\}$,  $\sum_k E_k^\dagger E_k = \mathbbm{1}$, be the Kraus representation of $\mathcal{C}$. The state after the loss of qubit $2$ is
\begin{eqnarray}
     & \tr_{\{2\}} \bigg[ \sum_k (\mathbbm{1}\otimes E_k \otimes\mathbbm{1})\, \rho\, (\mathbbm{1}\otimes E_k^\dagger \otimes\mathbbm{1}) \bigg]		\\
  =  & \tr_{\{2\}} \bigg[ \sum_k (\mathbbm{1}\otimes E_k^\dagger \otimes\mathbbm{1})  (\mathbbm{1}\otimes E_k \otimes\mathbbm{1})\, \rho \bigg]		\\
  =  & \tr_{\{2\}} \bigg[ (\mathbbm{1}\otimes \Big( \sum_k  E_k^\dagger E_k \Big) \otimes\mathbbm{1}) \, \rho \bigg]
  =    \tr_{\{2\}} \rho.
\end{eqnarray}
This shows that the remaining state after loss is entirely independent of the loss mechanism, as long as qubit 2 does not interact with other qubits as part of this loss mechanism. It is easy to see that this principle extends to the loss of many qubits.

\subsection{Pure states are sufficient for representing every state of the adaptive measurement\label{sec:Adaptive_Feedback_Measurement_Schemes_A}}

Initially, the measurement scheme starts with a pure permutationally-symmetric $n$-qubit state $\rho_\tn{in} = \ket{\Psi_n}\bra{\Psi_n}$, $\ket{\Psi_n} = \sum_{\nu=0}^n \psi_\nu \PiStateFull{\nu}{n}$ as input. Corollary \ref{theorem:Loss_can_be_posponed_to_the_End} implies that all measurement results are indistinguishable from a lossless experiment, where we successively measure the qubits $\{1,2,\ldots,m\}$, for $m = |M|$. Then, Remark \ref{Remark:Resulting_State_after_single_qubit_PVM} implies that the state after the first measurement can be written as 
\begin{eqnarray}\label{eq:state_after_Adaptive_Feedback_PVM}
  \rho_\ell &= \frac{1}{p_\ell} K_\ell\, \rho_\tn{in}\, K_\ell^\dagger
	\stackrel{\eqref{eq:Remark:Resulting_State_after_single_qubit_PVM-Eq1}}{=} 
	\frac{1}{p_\ell} \ket{\ell'}_1\bra{\ell'} \otimes \ket{\Psi_\ell}\bra{\Psi_\ell}    
	\quad \in 	\mathcal{B}(\mathscr{H}^{(1)}_{2}) \otimes \mathcal{B}(\mathscr{H}_{n}), \\
  \ket{\Psi_\ell} &= \frac{1}{\sqrt{n}} 	\sum_{\nu=0}^{n-1}
						\big(\sqrt{n-\nu}\Psi_\nu\kappa_{\ell,0} + \sqrt{\nu+1}\Psi_{\nu+1}\kappa_{\ell,1}\big)\PiStateFull{\nu}{n-1},
\end{eqnarray}
for $\ell$ the measurement outcome and $p_\ell = \sum_{\nu=0}^{n-1}  |\sqrt{n-\nu}\Psi_\nu\kappa_{\ell,0} + \sqrt{\nu+1}\Psi_{\nu+1}\kappa_{\ell,1}\big |^2$ the measurement probability. Hence, the state after the first measurement is pure, where the substring of the unmeasured qubits is in the permutationally-symmetric state $\ket{\Psi_\ell} \in \mathscr{H}_{n}$. 

After the first measurement, the processing unit selects a possibly different PVM to be applied to the next qubit of the unmeasured  permutationally-symmetric string $\ket{\Psi_\ell} \in \mathscr{H}_{n}$. The same argument as above can now be applied to the second measurement. By induction, it follows that at every state of the experiment, a pure state is sufficient for computing the probabilities of the measurement outcome.
Consequently, for a pure permutationally-symmetric qubit string measured by single-qubit PVMs, the density matrix formalism is dispensable.

\subsection{Exponential speedup in simulating adaptive-measurement schemes\label{sec:Adaptive_Feedback_Measurement_Schemes_B}}

\noindent%
By employing our mathematical framework, one can achieve an exponential speedup in simulating adaptive-measurement schemes based on sequential single-qubit PVMs. The naive approach suggests that, when an individual qubit of a permutationally-symmetric state $\rho \in \mathcal{B}\big(\mathscr{H}_{n+1}\big)$ is measured with a PVM, the permutation symmetry is broken and thus, the state after the measurement should to be represented as $\rho \in \mathcal{B}(\mathscr{H}_2^{\,\otimes n})$. In a classical computer simulation, storing $\rho$ in the memory can require exponentially many bits with respect to $n$. 
However, as we have shown in subsection \ref{sec:Adaptive_Feedback_Measurement_Schemes_A} above, the measured state is separable with the unmeasured qubit string in $\mathcal{B}\big(\mathscr{H}_{n}\big)$ and the measured qubit in $\mathcal{B}\big(\mathscr{H}_{2}\big)$. Hence, in a classical computer simulation, the state can be stored using $\mathcal{O}(n^2)$ bits, thereby requiring exponentially less memory compared to the naive approach. 

As the quantum channel $\mathcal{E}$ is unitary and the measurement is a PVM, the action of the combined PVM on a single qubit state can be computed in constant time. Furthermore, state \eqref{eq:state_after_Adaptive_Feedback_PVM} can be computed in time $\mathcal{O}(n)$\footnote{We use the well-established premise within computational complexity theory that elementary operations, such as addition, multiplication or the square root, can be computed in constant time.}. For simulating the experimental execution of an adaptive-measurement scheme, the resultant state after at most $n$ measurements has to be computed, requiring $\mathcal{O}(n^2)$ operations.

\subsection{Discussion}

One of the strengths of our approach is that the single-qubit PVMs need not be identical. Due to this versatility, our analysis also applies to multi-pass measurement schemes, for which permutationally-symmetric states are used as inputs. Examples include the spatial alignment of reference frames \cite{Rudolph&Grover:PhysRevLett:2003}, clock synchronization \cite{deBurgh&Bartlett:PhysRevA:2005}, and phase estimation \cite{Berry-Wiseman:PhysRevA:2009}. The spatial and temporal alignment of the reference frames is reduced to an adaptive multi-pass phase-estimation task \cite{Rudolph&Grover:PhysRevLett:2003,deBurgh&Bartlett:PhysRevA:2005}. 
Entangled permutationally-symmetric states are a key resource for multi-pass phase estimation, as they allow for increased precision in the presence of loss  \cite{Demkowicz-Dobrzanski:QuntInfSci:2010}.  
The applicability of our results to multi-pass measurement schemes using permutationally-symmetric states as inputs can be seen by the following argument. 
A multi-pass measurement scheme works in the same manner as an adaptive measurement scheme, except that each input qubit undergoes multiple passes through the channel $\mathcal{E}$ prior to the measurement. 
After all passes through $\mathcal{E}$ and a successful measurement of the input qubit, the number of passes is changed for the next input qubit.
For an input qubit, all passes though $\mathcal{E}$ and the subsequent projective measurement can be jointly represented by a PVM.

Entangled permutationally-symmetric qubit strings are a key resource for quantum-enhanced adaptive phase estimation and multi-pass phase measurements under lossy conditions. 
However, the usefulness of a permutationally-symmetric qubit string can be quite different under lossy conditions compared to the case of a lossless setup. For example, loss quickly destroys entanglement in the N00N state $(\PiStateFull{0}{n} + \PiStateFull{n}{n})/\sqrt{2}$ and thereby eliminates the advantage of N00N states over classical states \cite{Rubin&Kaushik:PhysRevA:2007}. However, entangled permutationally-symmetric states for phase estimation that are robust against loss are well known \cite{Optimal_Quantum_Phase_Estimation:PRL:2009,Maccone&DeCillis:PhysRevA:2009}.

Our results are mainly relevant for quantum-enhanced adaptive measurements and multi-pass schemes under lossy conditions, where entangled qubit strings are required to achieve an increased accuracy over classical schemes \cite{Quantum_phase_estimation_with_lossy_interferometers:PRA:2009,Demkowicz-Dobrzanski:QuntInfSci:2010}.
Unfortunately, generating the required entangled $n$-qubit permutationally-symmetric input states is generally beyond current technology for large $n$.

\section{Conclusions}

\noindent%
We have shown that the measurement probabilities of single-qubit POVMs are not affected by some qubit losses for permutationally-symmetric states. This loss-independence underscores the importance of permutationally-symmetric states for real-world applications.  When qubits of a permutationally-symmetric state are measured, the global permutation symmetry is broken. However, the unmeasured qubit substring remains permutationally-symmetric and therefore the loss-independence of later measurements of this substring persists (Corollary \ref{theorem:Loss_can_be_posponed_to_the_End}).

Moreover, our results are important for efficiently simulating adaptive-measurement schemes including loss. As has been shown, permutationally-symmetric states are the optimal inputs for phase estimation \cite{Quantum_phase_estimation_with_lossy_interferometers:PRA:2009}. Most adaptive phase-estimation schemes employ pure permutationally-symmetric states as inputs and projective measurements at the outputs to estimate properties of a unitary quantum channel  \cite{PhysRevLett.85.5098,PhysRevA.63.053804,Berry-Wiseman:PhysRevA:2009,QLearning:hentschel:PRL:2010,QLearning:hentschel:ITNG:2010}.
We have shown that, under these prerequisites, permutationally-symmetric states are sufficient to describe every state of the experiment thus allowing for a polynomial time simulation with respect to the number of input qubits. 

\ack
\noindent%
We thank D.\,Markham for valuable discussions and M.\,B.\,Ruskai for her critique of an earlier draft.
This project has been supported by \textit{i}CORE and NSERC. BCS is a CIFAR Fellow.

\newpage
\fontsize{8pt}{8pt}\selectfont

\end{document}